\documentclass[a4paper,superscriptaddress,twocolumn,prb,aps,floatfix,citeautoscript]{revtex4}
\usepackage{graphicx}
\usepackage{amsmath,amsbsy,amsfonts,mathrsfs}
\usepackage{color}

\newcommand{\D}{{\rm d}}

\newcommand{\tpci}{Theoretical and Physical Chemistry Institute, NHRF, Vas. 
Constantinou 48, GR-11635, Athens, Greece.}
\newcommand{\fu}{Institut f\"ur Theoretische Physik, Freie Universit\"at Berlin, 
Arnimallee 14, D-14195 Berlin, Germany.}
\newcommand{\fhi}{Fritz Haber Institute of the Max Planck Society,
Faradayweg 4-6, D-14195 Berlin, Germany.}
\newcommand{\kay}{School of Chemistry, The University of Edinburgh, 
Edinburgh EH9~3JJ, U.K.}
\newcommand{\lyon}{Laboratoire de Physique de la Mati\`ere Condens\'ee et Nanostructures, 
Universit\'e Lyon I,\\
CNRS, UMR 5586, Domaine Scientifique de la Doua, F-69622 Villeurbanne Cedex, France.}
\newcommand{\coimbra}{Centre for Computational Physics, Department of Physics, 
University of Coimbra, 3004-516 Coimbra, Portugal.}
\newcommand{\etsf}{European Theoretical Spectroscopy Facility.}

\begin{document}
\title
{Density-Matrix-Power Functional: Performance for Finite Systems and the 
Homogeneous Electron Gas}

\author{N.\,N. Lathiotakis}
\affiliation{\tpci}
\affiliation{\fu}
\affiliation{\etsf}

\author{S. Sharma}
\affiliation{\fhi}
\affiliation{\fu}
\affiliation{\etsf}

\author{J. K. Dewhurst}
\affiliation{\kay}

\author{F. Eich}
\affiliation{\fhi}
\affiliation{\fu}
\affiliation{\etsf}

\author{M.\,A.\,L. Marques}
\affiliation{\lyon}
\affiliation{\coimbra}
\affiliation{\etsf}

\author{E.\,K.\,U. Gross}
\affiliation{\fu}
\affiliation{\etsf}

\renewcommand{\vec}[1]{{\bf #1}}
\newcommand{\eref}[1]{(\ref{#1})}
\newcommand{\be}{\begin{equation}}
\newcommand{\ee}{\end{equation}}
\newcommand{\bea}{\begin{eqnarray}}
\newcommand{\eea}{\end{eqnarray}}
\newcommand{\nup}{n_{\uparrow}}
\newcommand{\ndown}{n_{\downarrow}}
\newcommand{\br}{{\bf r}}
\newcommand{\bx}{{\bf x}}

\begin{abstract}
An exchange correlation energy functional involving fractional power of the 
one-body reduced density matrix [Phys. Rev. B {\bf 78}, 201103 (2008)] is 
applied to finite systems and to the homogeneous electron gas. The performance 
of the functional is assessed 
for the correlation and atomization energies of the molecules contained in the
G2 set and for the correlation energy of the homogeneous electron gas. High 
accuracy is found for these two very different types of systems. 
\end{abstract}

\date{\today}
%\pacs{}

\maketitle
%%%%%%%%%%%%%%%%%%%%%
%%  INTRODUCTION   %%
%%%%%%%%%%%%%%%%%%%%%
For the past 40 years, density functional theory (DFT) developed into
one of the most successful theories in the study of the electronic
structure of atoms, molecules, and periodic solids. At its heart lies
the exchange-correlation (xc) functional, for which many 
approximations have been proposed. The simplest functionals, that
depend  only on the density (the local density
approximation -- LDA), or on the density and its gradients (the
generalized gradient approximation -- GGA), give a very
satisfactory description of many ground-state properties. However,
they still fail to reach chemical accuracy for some
important quantities like reaction or atomization energies.
To remedy this situation hybrid functionals were introduced, the first and most
widely used example being the B3LYP functional \cite{becke,lee}. 
This functional is able to reproduce experimental atomization energies within
about 10\% error. 
Although the atomization energies obtained using
B3LYP are in good agreement with experiments, the
absolute correlation energies, an accurate description of which can be thought
of as a test for the quality of any approximate functional, 
exhibit a sizeable error (up to 400\%) \cite{bench}. 
This is not a surprise since experimentally one normally measures energy
differences, and it is these quantities that functionals like B3LYP
are designed to reproduce.
Accurate correlation energies for finite systems can be obtained by going
beyond the DFT framework, for instance by using M{\o}ller-Plesset
second-order perturbation theory (MP2) or the coupled cluster method
with singles, doubles and perturbative triples [CCSD(T)]. However, these
methods are computationally too expensive to be applied to realistic
systems of ever growing complexity: bio-molecules, large clusters,
nano-devices to name but a few examples.

Recently, reduced density matrix functional theory (RDMFT) has
appeared as an alternative approach to handle complex systems. It has
shown great potential for improving upon DFT results for finite
systems.  
RDMFT uses the one-body reduced density matrix (1-RDM), $\gamma$, as the
basic variable \cite{lowdin,gilbert}. This quantity, for the ground state,
is determined through the minimization of the total energy functional, 
under the constraint that $\gamma$
is ensemble $N$-representable. 
The total energy as a functional of $\gamma$ can be expressed as
\begin{multline}
  \label{etot}
  E_v[\gamma] = \int\!\! \D^3r\!\!\int\!\! \D^3r'\; \delta({\bf r}-{\bf r}')
  \left[-\frac{\nabla^2}{2}\right]  \gamma({\bf r},{\bf r}') \\
  + \int\!\! \D^3r\; v({\bf r})  \rho({\bf r})
  +\frac{1}{2}\int\!\! \D^3r\!\!\int\!\! \D^3r'\; 
 \frac {\rho({\bf r})\rho({\bf r}')}{|{\bf r}-{\bf r}'|}
  + E_{\rm xc}[\gamma]
  \,,
\end{multline}
where, $\rho({\bf r})$ ---the electron density--- is the diagonal of the 1-RDM 
and $v({\bf r})$ is the external potential. 
The first two terms in 
Eq.~(\ref{etot}) are the kinetic and external potential energies. 
The electron-electron interaction can be cast in the last two terms, the 
first being the Coulomb repulsion energy and $E_{\rm xc}$  the
exchange-correlation (xc) energy functional.
In principle, Gilbert's \cite{gilbert} generalization of the Hohenberg-Kohn
theorem to the 1-RDM guarantees the existence of a functional $E_v[\gamma]$ 
whose minimum yields the exact $\gamma$ and the exact ground-state energy of the 
systems characterized by the external potential $v({\bf r})$. In practice, however,
the xc energy is an unknown functional of the 1-RDM and needs to be
approximated. In the last
years, a plethora of approximate functionals have been introduced.
\cite{kollmar,kios1,kios2,pernal1,csanyi,nekjellium,csgoe,gritsenko,piris,pade}
An assessment of the relative performance of these functionals for a large
set of atoms and molecules reveals that the
so called BBC3 \cite{gritsenko} and PNOF0 \cite{piris} functionals
yield results for molecular systems, with an accuracy comparable
to the MP2 method \cite{piris,piris_os,piris_JTCC,bench,leiva,pernalip}.

The situation for extended systems is somewhat more complicated. When
applied to the simplest system, the homogeneous electron gas, these
functionals lead to rather inaccurate correlation energies \cite{nekjellium}. 
Moreover, for periodic solids, the aforementioned functionals fail to reproduce
the fundamental gaps for insulators
\cite{sharma} (band as well as Mott). A new functional was introduced by 
Sharma \emph{et al.} to solve this problem.\cite{sharma} It reads
\begin{equation}
  \label{sdlg}
   E_{\rm xc}[\gamma]=
  -\frac{1}{2}\int\!\! \D^3r\!\!\int\!\! \D^3r'\;
  \frac{|\gamma^{\alpha}({\bf r},{\bf r}')|^2}{|{\bf r}-{\bf r}'|}
  \,,
\end{equation}
where, $\gamma^{\alpha}$ indicates the power used in the operator
sense i.e., 
\begin{equation}
\label{gpa}
\gamma^{\alpha}({\bf r},{\bf r}')=\sum_j (n_j)^{\alpha} \phi^*_j({\bf r}')
\phi_j({\bf r})
\end{equation}
Here $\phi_j({\bf r})$ denote the natural orbitals and $n_j$ their occupation
numbers. The functional in Eq. (\ref{sdlg}) was shown to perform very well for 
solids \cite{sharma}. As it involves the power of the density matrix we refer to
it as "power functional" in the following. The power $\alpha$ lies in the 
interval $1/2 \le \alpha \le 1$. In the limit $\alpha=1$, minimization of the 
total energy functional (\ref{etot}) yields the Hartree-Fock energy, while the
case $\alpha=1/2$ corresponds to the M\"uller functional which tends to over 
correlate \cite{nekjellium,bench}. Hence the power $\alpha$ plays the similar 
role as the mixing parameter in the hybrid functionals of the DFT.

The situation as it stands is that  most of the existing xc functionals 
of RDMFT are
designed for finite systems and perform quite poorly for solids and
the HEG.  The power-functional, on the other hand, is designed for the
case of solids, but has not yet been applied to the HEG or finite
systems. The objective of the present work is to fill this gap, in
the pursuit of a functional form that works equally well for both
finite and extended systems.

%%%%%%%%%%%%%%%%%%%%%%%%%%%%%%%%
% Results and discussion: HEG
%%%%%%%%%%%%%%%%%%%%%%%%%%%%%%%%

\begin{figure}[ht]
\centerline{\includegraphics[width=0.7\columnwidth,angle=-90]{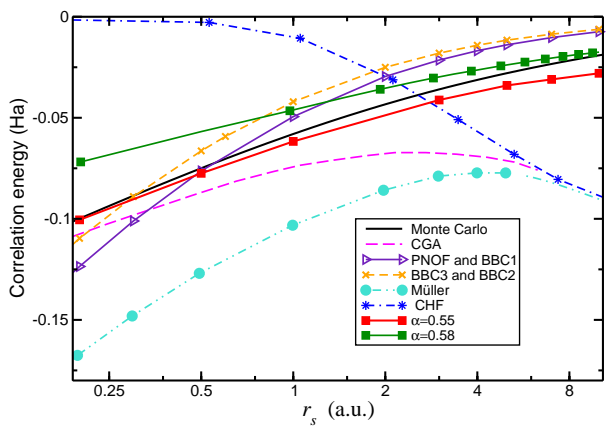}}
\caption{(Color online) Correlation energy as a function of
  the Wigner-Seitz radius for the homogeneous electron gas. RDMFT results
  are obtained using various approximations to the xc functional.
  Monte Carlo results are taken from Ref.~~[\onlinecite{OB}]. }
\label{jel}
\end{figure}

It is difficult to overstate the importance of the HEG in the
development of many-body theories. It is not only used as a
benchmark, but also acts as a reference system for the design
of xc functionals. Within DFT, the LDA is perhaps the most
successful example of this. Furthermore, results for the HEG give us
valuable indications on how the theory will perform especially for
metallic systems.

With this in mind, we first compare the relative performance of various
RDMFT functionals in reproducing the correlation energy of the HEG of
various densities. Fig. \ref{jel} is a plot of the correlation
energy versus the density parameter $r_s$ for a variety of approximate
functionals compared to exact Monte-Carlo values.\cite{OB} Clearly,
the M\"{u}ller, \cite{mueller} CHF \cite{csanyi} and CGA \cite{csgoe}
functionals perform poorly over the whole range of $r_s$. The BBC
\cite{gritsenko} and PNOF\cite{piris} functionals are more reasonable but still
far from the true result.
Encouragingly, we find that for $\alpha$ between $0.55$ and $0.58$,
the power-functional lies very close to the Monte Carlo results and
possesses a good low density limit, making it one of the best 1-RDM
functionals for the HEG.

%%%%%%%%%%%%%%%%%%%%%
%% RnD: Ec-molecules
%%%%%%%%%%%%%%%%%%%%%
Since the power-functional performs well for the HEG at small values
of $r_s$ and for periodic solids \cite{sharma}, it is worthwhile to
investigate its behavior for finite systems. First we performed a free
optimization of the parameter $\alpha$ using a set of 54 molecules and
radicals. These molecules form a subset of the G2 test set of
molecules.~\cite{g2set1,g2set2} For this optimization two different
basis sets are employed, namely the 6-31G* and the cc-pVDZ.  The
optimal value of $\alpha$ that minimizes the error in the correlation
energy for this set of molecules is 0.578. This value is essentially the same 
for both kinds of basis set used in the present work. It is interesting to
note from Fig. \ref{jel} that the value of $\alpha$ which best
reproduces the Monte-Carlo data for $r_s$ of interest for most atoms
and molecules is about 0.55. The optimal value obtained for the set of 
molecules, $\alpha=0.578$, is quite close to this.

\begin{figure}
\vspace{0.5cm}
\includegraphics[width=0.7\columnwidth,angle=-90]{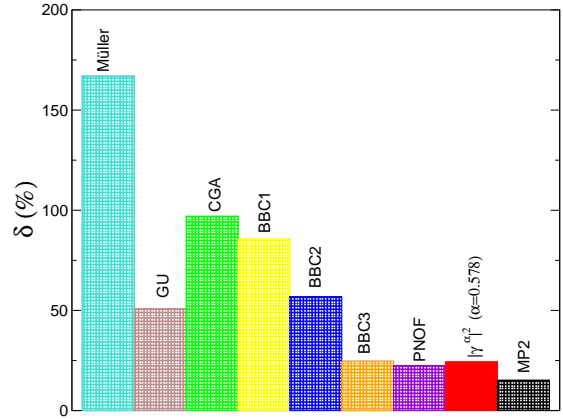}
\caption{(Color online) Percentage deviation of the correlation energy, obtained
using various 1-RDM functionals, from the exact CCSD(T) results.}
\label{ec-mol}
\end{figure}

Having determined the optimal value for the parameter $\alpha$, we
compare this approach with different RDMFT functionals.  Fig.
\ref{ec-mol} is the plot of relative error in the correlation energy
($\delta$) given by
\begin{equation}
  \label{delta}
  \delta = 100 \sqrt{\sum \left[(E_{\rm c} - 
  E_{\rm c}^{\rm (ref)})/E_{\rm c}^{\rm (ref)}\right]^2/N_{\rm mol}}\,. 
\end{equation}
where $E_{\rm c}$ is the correlation energy obtained with the method
under consideration, $E_{\rm c}^{\rm (ref)}$ the reference correlation
energy which is obtained with the CCSD(T) method, and $N_{\rm mol}$
the number of systems included in the calculation.  It is apparent
from Fig. \ref{ec-mol} that the power-functional is very good in
determining the correlation energy of finite systems. In fact, we find
that this very simple functional with $\alpha$=0.578 (same value of $\alpha$ was
used for the full G2 set) is of similar
quality as the BBC3 or the PNOF0 which are to-date the best RDMFT
functionals as far as finite systems are concerned.
\cite{piris,piris_os,piris_JTCC,bench,leiva,pernalip}

%%%%%%%%%%%%%%%%%%%%%
%% RnD: AE-molecules
%%%%%%%%%%%%%%%%%%%%%
\begin{figure}
\vspace{0.5cm}
\includegraphics[width=0.75\columnwidth,angle=-90]{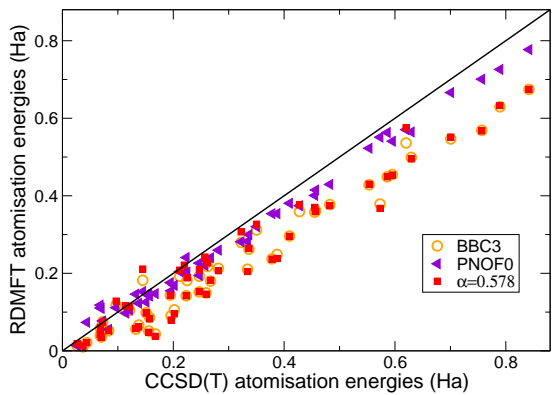}
\caption{(Color online) Atomization energies for the G2 set of molecules
  calculated using the BBC3, PNOF0 and power-functional vs the CCSD(T)
reference atomization energies.}
\label{ae_ae}
\end{figure}

The accurate determination of atomization energies is important for
the calculation of the energetics of any chemical reaction. Hence this
important quantity also acts as a test for the quality of an
approximate functional. In Fig.~\ref{ae_ae} we plot the atomization
energies of molecules of the entire G2 set obtained with various
approximate functionals within RDMFT versus the reference value
determined using the CCSD(T) method.

It is clear from Ref.~\onlinecite{bench} and Fig.~\ref{ec-mol}
that BBC3, PNOF0 and the power-functional are the the most accurate among
the xc functionals we considered, hence
in Fig.  \ref{ae_ae}, we only compare these three functionals.  
It is clearly visible that the power-functional is comparable in accuracy to
the BBC3 functional, while PNOF0 is slightly better than the two.

%%%%%%%%%%%%%%%%%%%%%
%%  H2 dissociation 
%%%%%%%%%%%%%%%%%%%%%
 
\begin{figure}
\vspace{0.5cm}
\includegraphics[width=0.8\columnwidth,angle=-90]{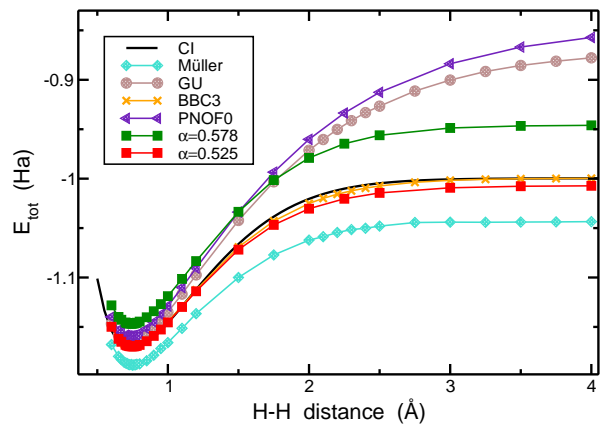}
\caption{ (Color online) Total energy of the H$_2$ molecule (in a.u.) vs
  the distance (in \AA) between the two hydrogen atoms. 
  RDMFT results are obtained using various approximations to the xc functional. 
  For reference the configuration interaction results are also presented 
  (black line).}
\label{H2} 
\end{figure}

The successful prediction of properties of molecules at equilibrium does not 
necessarily imply a good performance for strongly distorted molecules, the
dissociation limit  being one such example. For a stretched molecule, 
not only the total energy has to be equal to the sum of the energies of 
the fragments but also the occupations of the natural orbitals have to be
correct.  The simplest example is perhaps the H$_2$ molecule. 
If we take the hydrogen atoms far apart, the total energy should go to -1 a.u.
and the four occupied spin-orbitals made from the Hydrogen 1$s$ states, 
have to be occupied by half an electron each. 
Many DFT and RDMFT functionals fail to reproduce the correct dissociation
of this system\cite{baerends,grit}.

In Fig.~\ref{H2}, we plot the H$_2$ dissociation curve obtained using
various 1-RDM functionals, together with the exact curve obtained
through a full configuration-interaction calculation.  The BBC3
functional is designed with the dissociation limit in mind and
it yields the most accurate results for the H2 dissociation. The failure 
of PNOF0 and GU functionals is two fold; as the distance between the two H
atoms increases, the energy deviates strongly from the exact value.
Second, at a sufficiently large distance between the two H atom (6\,
\AA) two of the bonding orbitals still have occupancy 0.86 and the
other two 0.14. Both the M\"uller and the power-functional with
$\alpha=0.578$ give the correct occupancy of $\sim$0.5 for all four
bonding spin-orbitals, but the dissociation energy is underestimated
by the former of these functionals and overestimated by the later. 
If, on the other hand, the value of $\alpha$ is
changed to 0.525 the power-functional describes the H2 dissociation
curve accurately, with an accuracy comparable to the BBC3 functional.

In all the examples studied, it is clear that the simple form of the
power-functional suffices to obtain very good electronic properties.
However, we are faced with a problem: the optimal value for $\alpha$
varies from one kind of system to another. In fact, we find
$\alpha=0.65$ for solids, $\alpha=0.55$ for the HEG, $\alpha=0.578$
for molecules at equilibrium, and $\alpha=0.525$ for stretched H$_2$.
Although one can use different values of $\alpha$ for different
materials, it would be desirable to have a unique method to determine the
system dependent value of $\alpha$. In this regard, one could make $\alpha$
itself a functional of the 1-RDM and optimize it in as \emph{ab-initio} manner 
for each system. Many other improvements of the power-functional are also
conceivable; for example it could be a
basis for sophisticated corrections like those of Gristenko \emph{et. al}
\cite{gritsenko} and/or removal of self-interaction terms. How these
modifications affect the quality of the power-functional will be the
subject of future studies.

%%%%%%%%%%%%%%%%%%%%%
%%   Conclusions
%%%%%%%%%%%%%%%%%%%%%

In summary, we have used the recently proposed power-functional to
calculate the correlation energy of the HEG, G2 test set of
molecules, and stretched H$_2$. For the case of molecules, we also
determined atomization energies. Our results show that the
power-functional, originally proposed for solids, also performs very
well for the HEG and finite systems.  However, the optimal value
of $\alpha$ for all three cases is different.  The
encouraging results of the present work point to the fact that this
family of approximations is an interesting path for the future
development of approaches, within RDMFT, to accurately describe
electronic correlations.

%%%%%%%%%%%%%%%%%%%%%
%% Acknowledgements
%%%%%%%%%%%%%%%%%%%%%
This work was supported by the Deutsche Forschungsgemeinshaft (SPP 1145), and 
by the EC Network of Excellence NANOQUANTA (NMP4-CT-2004-500198). Part of the 
calculations were performed at the Laborat\'orio de Computa\c{c}\~ao 
Avan\c{c}ada of the University of Coimbra.

%\bibliography{alpha}

\end{document}